\documentclass[9pt,twocolumn,twoside]{osajnl}

\journal{ol} 

\setboolean{shortarticle}{true} 

\title{Photonic simulation of giant atom decay}
\author[1,2,*]{Stefano Longhi}
\affil[1]{Dipartimento di Fisica, Politecnico di Milano, Piazza L. da Vinci 32, I-20133 Milano, Italy}
\affil[2]{IFISC (UIB-CSIC), Instituto de Fisica Interdisciplinar y Sistemas Complejos - Palma de Mallorca, Spain}
\affil[*]{Corresponding author: stefano.longhi@polimi.it}

\setboolean{displaycopyright}{true}

\dates{Compiled \today}

 \ociscodes{(130.3120) Integrated optics devices;  (270.5580)   Quantum electrodynamics; (080.1238) Array waveguide devices}
\doi{\url{http://dx.doi.org/10.1364/ol.XX.XXXXXX}}
\begin{abstract}
Spontaneous emission of an excited  atom in a featureless continuum of electromagnetic modes is a fundamental process
in quantum electrodynamics associated  with an exponential decay of the quantum emitter to its ground state accompanied
by an irreversible emission of a photon. However, such a simple scenario is deeply modified when considering a $^{\prime}$giant$^{\prime}$ atom, i.e an atom whose
dimension is larger than the wavelength of the emitted photon. In such an unconventional regime, non-Markovian effects and strong deviations from an exponential decay 
are observed owing to interference effects arising from non-local light-atom coupling. Here we suggest a photonic simulation of non-Markovian giant atom decay, based on light escape dynamics in an optical waveguide  non-locally-coupled to a waveguide lattice. Major effects such as 
non-exponential  decay, enhancement or slowing down of the decay, and formation of atom-field dark states can be emulated in this system.
 \end{abstract}
\setboolean{displaycopyright}{true}
\begin{document}
\maketitle
\thispagestyle{fancy}
\ifthenelse{\boolean{shortarticle}}{\abscontent}{} 

{\it Introduction.}  
In traditional light-matter interaction, atoms are usually considered point-like quantum emitters 
and atom-field coupling is local in the dipole approximation \cite{r1}. For an excited atom emitting in a featureless continuum of electromagnetic modes in the vacuum state,
the resulting process of spontaneous emission is well described by an exponential decay of atomic excitation to its ground state associated to an irreversible emission of a photon, according to the 
Weisskopf-Wigner theory \cite{r2}.This picture, however, is challenged when considering spontaneous decay of $^{\prime}$giant$^{\prime}$ atoms (GAs), i.e. artificial quantum emitters whose dimension is larger than the wavelength of the emitted photon \cite{r3,r4,r5,r6,r7,r8,r9,r10,r11,r12,r13}. In this case the time it takes for light to pass a single atom cannot be neglected, giving rise to strong non-Markovian dynamics at the single atom level even thought the atom-field coupling is weak. Major effects arising from the time-delay dynamics are  strong deviations from exponential decay, decoherence-free atomic interactions, chiral emission, and frequency-dependent Lamb shifts \cite{r3,r7,r8,r9,r10}.
Recent experimental implementations of GAs are based on superconducting qubits,  coupled either to surface acoustic waves \cite{r3,r4,r6,r8} or to microwaves in a waveguide at distant positions \cite{r10}. Quantum simulations of non-local light-matter coupling Hamiltonians have been also suggested using ultracold atoms in in dynamical
state-dependent optical lattices \cite{r9}. A simple way to implement GA non-Markovian decay is to couple a point-like emitter to distant bath positions, inducing strong interference
effects among emission occurring at different time instants.
 In a different area of research, lattices of evanescently coupled optical waveguides
probed with either classical or nonclassical states of light
have provided, over the past two decades, a useful laboratory
tool for simulating a wealth of coherent quantum phenomena in the matter \cite{r14,r15,r16,r17,r18}. Waveguide arrays
 can effectively emulate light-matter coupling Hamiltonians, and have been harnessed
to visualize in optical settings phenomena like Zeno and anti-Zeno dynamics \cite{r19,r20,r21}, non-exponential quantum decay at long times \cite{r22}, 
 the ultra strong coupling regime of light-matter interaction \cite{r23,r24,r25,r26}, dark states \cite{r27,r28,r29}, decoherence of optical Schr\"odinger cat states \cite{r30}, Majorana dynamics \cite{r30b}, and 
 decay in multi-dimensional space \cite{r30c}.\\
 In this Letter we suggest a photonic simulation of non-Markovian decay dynamics of a giant atom, where an optical waveguide (emulating a point-like emitter) is side-coupled to distant points of a waveguide lattice (the continuum). In the single excitation sector of Hilbert space, photon escape along the waveguide exactly reproduces the decay dynamics of a point-like quantum emitter weakly coupled to a featureless continuum of bosonic modes. Depending on the discretized distance between waveguide-lattice contact points, different non-Markovian regimes can be observed, such as enhancement as well as fully or partial suppression of the decay owing to the appearance of bound states in the continuum.\\
 \par
 {\it Photonic system and basic model.} Let us consider the photonic system schematically depicted in Fig.1(a), consisting of a straight dielectric optical waveguide W which is side-coupled by evanescent field to an infinitely-extended one-dimensional waveguide lattice. The lattice is bent in the $(x,y)$ plane to form a rotated U-shaped path, so as W is effectively coupled to distant sites of the lattice near $n=0$ and $n=n_0$, as shown in Fig.1(a). Photon propagation in the system is described by the tight-binding Hamiltonian (see e.g. \cite{r30,r31})
 \begin{equation}
 \hat{H}=\hat{H}_a+\hat{H}_{b}+\hat{H}_{int}
 \end{equation}
 \begin{figure}[htb]
\centerline{\includegraphics[width=8.7cm]{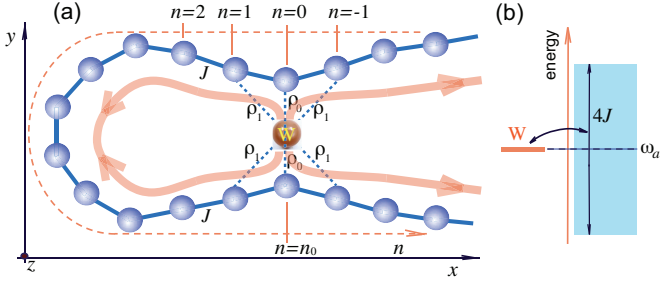}} \caption{ \small
(Color online) (a) Schematic of the integrated photonic system emulating GA decay dynamics. A waveguide W is side-coupled to a waveguide lattice, which is bent in the $(x,y)$ plane to form a rotated U-shaped path. Evanescent field coupling occurs between waveguide W and the waveguides in the lattice in the neighborhood of the distant sites $n=0$ and $n=n_0$ (dashed bonds). $J$ is the coupling constant between adjacent waveguides is the lattice (solid bonds), and $\rho_l$ are the W-lattice coupling constants (with $\rho_l$ rapidly decreasing as $l$ is increased). All waveguides are straight along the propagation direction $z$ (optical axis). (b) Schematic of a  quantum emitter W decaying into a bath of bosonic oscillators forming a tight-binding energy band of width $4J$. In (a) the arrows indicate the 
four routes of light escaping from waveguide W into the lattice. The closed loop on the left side provides delayed back-excitation into the waveguide W, which is responsible for non-Marokian effects.}
\end{figure} 
 where $\hat{H}_a= \omega_a \hat{a}^{\dag} \hat{a}$ is the Hamiltonian of the photon field in waveguide W,  $\hat{H}_b=  J \sum_n ( \hat{b}^{\dag}_n \hat{b}_{n+1}+ H.c.)$ is the tight-binding Hamiltonian of the photonic modes  of the lattice, and $\hat{H}_{int}= \sum_{l=0, \pm 1, \pm2, ..}  \rho_{|l|} ( \hat{a}^{\dag} \hat{b}_{l}+\hat{a}^{\dag} \hat{b}_{l+n_0}+H.c.)$ is the Hamiltonian describing evanescent mode coupling of waveguide W with the lattice. In the above equations, $\hat{a}^{\dag}$ and $\hat{b}^{\dag}_l$ are the bosonic creation operators of photons in waveguide W and in the $l$-th waveguide of the lattice, respectively, $J$ is the coupling constant between adjacent waveguides in the lattice, $\omega_a$ is the propagation constant shift of waveguide W from the propagation constant of the lattice waveguides ($|\omega_a| \ll J$), and $\rho_l$ are the coupling constants between W and lattice waveguides, as shown by the dotted bonds in Fig.1(a). Since the coupling constant is a nearly-exponential decaying function of waveguide spacing \cite{r32}, we assume that $\rho_l$ takes its largest value at $l=0$, rapidly decaying toward zero as $l$ is increaased. Moreover, we assume the weak coupling regime $\rho_0 \ll J$, which is realized by closely-spacing the waveguides in the array. The bosonic creation/destruction operators of photons in the various waveguides of the system satisfy the usually commutation relations $ [\hat{a}, \hat{a}^{\dag}]=1$, $[ \hat{b}_l, \hat{b}^{\dag}_n]= \delta_{n,l}$, etc.  To describe photon escape dynamics in waveguide W, it is worth switching from Wannier to Bloch basis representation \cite{r30,r33} by introducing the bosonic operators $\hat{c}(k) \equiv (1/ \sqrt{2 \pi}) \sum_n \hat{b}_n \exp(-ik n)$ for Bloch modes, where $-\pi \leq k < \pi$ is the Bloch wave number and the bosonic commutation relations 
 $  [ \hat{c}(k), \hat{c}^{\dag} (k^{\prime}) ] = \delta (k-k^{\prime})$  
 hold. In such a representation, the full Hamiltonian of the photon field is given by Eq.(1) with
\begin{eqnarray}
\hat{H}_a & = & \omega_a \hat{a}^{\dag} \hat{a} \nonumber \\
\hat{H}_b & = & \int_{-\pi}^{\pi} dk \; \omega(k) \hat{c}^{\dag}(k) \hat{c}(k) \\
\hat{H}_{int} & = & \int_{-\pi}^{\pi} dk \left\{ G(k) \hat{a}^{\dag} \hat{c}(k) +H.c.  \right\} \nonumber
\end{eqnarray}
where 
\begin{equation}
\omega(k) \equiv 2 J \cos k
\end{equation}
is the dispersion relation of the tight-binding energy band and $G(k)=G_0(k) (1+ \exp(ik n_0))$ is the spectral coupling function, with
\begin{equation}
G_0(k) \equiv \frac{1}{\sqrt{2 \pi}} \sum_{l=0, \pm 1, \pm 2,...} \rho_{|l|} \exp(ikl).
\end{equation}
For the following analysis, we should distinguish two different regimes of waveguide-lattice coupling: the {\it short-range} coupling regime, corresponding to  $\rho_l=0$ for $|l|>0$, and the {\it long-range} coupling regime, corresponding to nonvanishing $\rho_l $ for some $ l \neq 0 $. Note that in the short-range coupling limit the spectral coupling function $G_0(k)$ is homogeneous, independent of $k$.\\
{\it Deacy dynamics.}
The Hamiltonian $\hat{H}$ of the photon field basically describes a bosonic harmonic oscillator of frequency $\omega_a$ weakly coupled to a bath of  harmonic oscillators of frequency $\omega(k)$. In the single excitation sector of Hilbert space and assuming the bath initially in the vacuum state and at zero temperature,  
the photonic system effectively emulates the decay dynamics of a point-like quantum emitter W non-locally coupled to a tight-binding continuum of bosonic oscillators [Fig.1(b)]. In fact, in the single excitation sector the state vector $ | \psi(z) \rangle$ of the photon field  can be written as
\begin{equation}
 | \psi(z) \rangle = a(z) \hat{a}^{\dag} | 0 \rangle + \int dk c(k,z) \hat{c}^{\dag}(k) |0 \rangle
\end{equation}
where the amplitude probabilities $a(z)$ and $c(k,z)$ satisfy the coupled equations
\begin{eqnarray}
i \frac{ da}{dz} & = & \omega_a a+\int dk G_0(k) \left( 1+ \exp(ikn_0) \right) c(k,z) \\
i \frac{\partial c}{\partial z} & = & \omega(k) c(k,z)+G^*_0(k) \left( 1+ \exp(-i k n_0) \right) a
\end{eqnarray}
with $a(z=0)=1$ and $c(k,z=0)=0$ for initial excitation of the system with one photon in waveguide W. The space-to-time relation $z=ct$ holds ($z=t$ in units of the speed of light $c$). The evolution of $a(z)$ effectively emulates the spontaneous emission decay process of a quantum emitter, and can be calculated by standard Laplace transform methods \cite{r22,r33,r34}. One obtains
\begin{equation}
a(z)=\frac{1}{2 \pi } \int_{-i \infty+o^+}^{i \infty+o^+} ds \frac{\exp(sz)}{is - \omega_a-\Sigma(s)}
\end{equation}
where
\begin{equation}
\Sigma(s) \equiv \int_{-\pi}^{\pi} dk \frac{2 |G_0(k)|^2 (1+ \cos(k n_0))}{is-\omega(k)}
\end{equation}
is the self-energy.\\
\par
{\it Delay-induced non-Markovian dynamics.} To highlight delay effects in the decay dynamics of $a(z)$ due to the non-local coupling of W with the lattice, let us focus our attention to the most interesting case $\omega_a=0$. Since the coupling is weak ($ \rho_0 \ll J$, i.e. $G_0 \rightarrow 0$), 
the main contribution to the integral on the right hand side of Eq.(8) comes from $s \sim 0$ as $s$ spans the imaginary axis. In fact, in the $G_0 \rightarrow 0$ limit the function under the sign of the integral shows a pole. We can thus consider an approximant to the self-energy $\Sigma(s)$ around $s=0$. From an inspection of Eq.(9), it follows that for $s \sim 0$ the main contribution to the integral on the right hand side of Eq.(9) comes when $k$ crosses the two points $k= \pm \pi/2$, where $\omega(k)=0$. Linearizing the dispersion curve $\omega(k)$ around $k= \pm \pi /2$, the integral on the right hand side of Eq.(9) can be readily computed by an asymptotic analysis. Provided that $G_0(k)$ varies slowly over $k$, which is strictly valid in the short-range coupling limit, one obtains 
\begin{equation}
\Sigma(s) \simeq -i \gamma \left[   1+ \exp(-i \phi -n_0 s /v_g) \right]
\end{equation}
where $v_g \equiv 2 J$ is the group velocity in the waveguide lattice at the band center, $ \phi \equiv \pi n_0/2$ is a quantized phase,  and  
\begin{eqnarray}
\gamma & \equiv & \frac{4 \pi} { v_g} |G_0 ( \pi/2)|^2=\frac{1}{J} \left| \sum_l \rho_{|l|} \exp(i l \pi/2) \right|^2 \nonumber \\
& = &\frac{1}{J} \left( \rho_0-2 \rho_2+2 \rho_4-2 \rho_6+...  \right)^2
\end{eqnarray}
is the effective decay rate. From Eqs.(8) and (10), it can be readily shown that $a(z)$ satisfies the following position-delayed differential equation
\begin{equation}
 \frac{da}{dz}=- \gamma a(z)- \gamma \exp(-i \phi) a(z- \tau) \theta(z-\tau)  
\end{equation}
where $\theta(z)$ is the Heaviside step function and $\tau \equiv n_0 / v_g$ is the propagation distance required by an excitation in the bath to travel from $n=0$ to $n=n_0$ sites. 
The decay dynamics described by Eq.(12) is typical of GAs coupled to a featureless continuum, showing strong deviations from an exponential decay and non-Markovian effects when the time delay $\tau$ becomes comparable or larger than the lifetime $1 / \gamma$ \cite{r3,r5,r8,r11}. We note that similar non-Markovian effects arising from delay effects can be observed for point-like emitters placed in from of a mirror \cite{r35,r36,r37}, as well as for atoms with resonances near a photonic band edge without delay \cite{referee}.  In our photonic system the last term on the right-hand side of Eq.(12) basically describes back-excitation of waveguide W from the light earlier decayed into the lattice after traveling along the closed loop indicated by the solid arrows in Fig.1(a). The solution to Eq.(12), which provides an approximation to the exact result given by Eq.(8), reads \cite{r36}
\begin{eqnarray}
a(z) & = & \exp(- \gamma z) \sum_{n=0,1,2,...}  \frac{  (-\gamma)^n \exp(-i n\phi+ n\gamma \tau) }{n !} \times  \nonumber \\
 & \times & (z-n \tau)^n \theta(z-n \tau).
\end{eqnarray}
\begin{figure}[htb]
\centerline{\includegraphics[width=8.7cm]{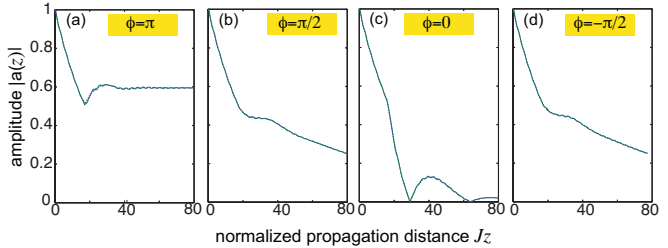}} \caption{ \small
(Color online) Decay dynamics (behavior of $|a(z)|$ versus normalized propagation distance $Jz$) in the short-range coupling for $ \rho_0 / J=0.2$ ($\rho_l=0$ for $l \geq 1$) and for a few values of site distance $n_0$, corresponding to different values of the phase $\phi=n_0 \pi /2$: (a) $n_0=34$ ($\phi= \pi$), (b) $n_0=33$ ($ \phi= \pi/2$), (c) $n_0=32$ ($\phi=0$), and (d) $n_0=31$ ($ \phi= - \pi/2$). Solid lines refer to the exact behavior of the decay dynamics [Eq.(8)], obtained from numerical solutions of coupled equations (6) and (7), while the dashed curves (almost overlapped with the solid ones) refer to the approximate decay behavior as given by Eq.(13). In (a) the decay is limited owing to the existence of a bound state in the continuum.}
\end{figure} 
\begin{figure}[htb]
\centerline{\includegraphics[width=8.7cm]{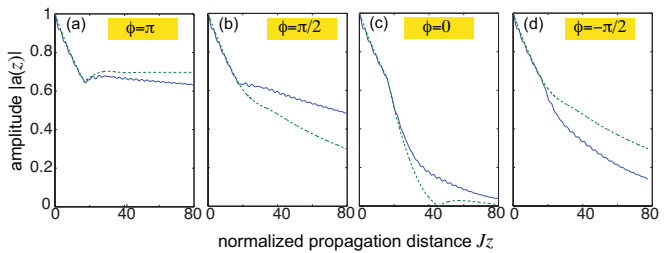}} \caption{ \small
(Color online) Same as Fig.2, but in the long-range coupling regime  ($ \rho_0 / J=0.2$, $\rho_1 / J=0.04$, $\rho_2 / J =0.02$, $\rho_l=0$ for $l \geq 3$). }
\end{figure} 
Note that in the interval $0<z<\tau$ the decay law is exponential with a decay rate $\gamma$ given by Eq.(11), while deviations from an exponential decay and non-Markovian effects are observed for $z> \tau$ due to the delayed back-excitation of waveguide W. Such effects are visible provided that the delay $\tau= n_0 /(2J)$ is comparable or larger than the lifetime $1 / \gamma$, and we thus limit to consider such a region of parameter space. We compare the results obtained from the exact decay dynamics [Eq.(8)] and the approximate dynamics [Eq.(13)] based on the differential-delayed equation (12).
Typical numerical results are shown in Figs.2 and 3 for the short-range and long-range coupling regimes, respectively. As expected, the predictions based on Eq.(12) are exact in the short range coupling limit (Fig.2), where the spectral coupling $G_0(k)$ is independent of $k$, while deviations are clearly observed in the long coupling regime (FIg.3). In both regimes,
three different dynamical behaviors are clearly observed, depending on the value of the quantized phase $\phi$. For an odd value of $n_0$, corresponding to a phase $\phi= \pm \pi/2$, after the initial exponential decay with a rate $\gamma$, the decay slows down [panels (b) and (d) in Figs.2 and 3], while for $n_0$ integer multiple of 4, corresponding to a phase $\phi=0$, the decay  becomes faster [panels (c) of Figs.2 and 3]. This means that the delayed feedback dynamics induced by the non-local coupling can emulate either sub-radiant and super-radiant spontaneous emission dynamics. For $n_0=2(2l+1)$ with $l$ integer, corresponding to a phase $\phi=\pi$, in the short-range coupling regime [Fig.2(a)] the decay is limited. This result stems  from the existence of a bound state in the continuum \cite{r37}, which is a stationary solution to the differential-delayed equation (12). It should be noted that the existence of a bound state in the continuum is an {\em exact} result for short-range coupling, i.e. its existence is ensured beyond the approximations leading to the differential-delayed equation (12). In fact, in the Wannier-basis representation and provided that  $\omega_a=0$, $\rho_l=0$ for $l \geq 1$, it can be readily shown that the Hamiltonian $\hat{H}$ in the single excitation sector shows an {\em exact} zero-energy eigenstate which localizes photons between sites $n=0$ and $n=n_0$ of the lattice, given by
\begin{eqnarray}
a= \mathcal{N} \; , \; \; b_l=0 \; \; (l<0 \; {\rm and} \;  l>n_0) \nonumber \\
 b_l=- \mathcal{N} (\rho_0 /J) \sin (  \pi l /2) \; \; (0 \leq l \leq n_0) \nonumber
\end{eqnarray}
where $\mathcal{N}$ is a normalization constant. The existence of such a bound state in the continuum arises from a destructive interference effect typical in photonic systems \cite{r33,r38,r39}.
In the long-coupling regime, the predictions based on Eq.(12) show consistent deviations from the full numerical simulations (Fig.3), mainly because $G_0(k)$ is not homogeneous in the whole Brilloujne zone  when $\rho_l$ is non-vanishing for some $l \geq 1$. From a physical viewpoint, the long-range couplings $\rho_1$, $\rho_2$, ... introduce additional delays $\tau_1= (n_0 \pm 1)/v_g$, $\tau_2= (n_0 \pm 2)/v_g$, ..., making the back action of such terms into the decay dynamics of $a(z)$ more involved, requiring to add additional delay terms $a(z-\tau_1)$, $a(z-\tau_2)$, ... in Eq.(12). The main impact of such multiple delayed feedback is that the decay is strongly slowed down but not fully suppressed for $\phi=\pi$ [Fig.3(a)], indicating that for non-negligible long-range couplings the bound state in the continuum actually becomes a quasi-bound state (a resonance).\\
The non-Markovian dynamics predicted by the above analysis should be feasible for an experimental observation using optical waveguide lattices realized by the femtosecond laser writing technology \cite{r16,r22,r24,r32,r39}. For example, let us assume passive optical waveguides manufactured in fused silica and probed in the red ($\lambda=633$ nm); the coupling constant $\kappa$ between two waveguides spaced by a distance $x$ follows an almost exponential law $\kappa=A \exp(-\sigma x)$ with $A \simeq 13.89 \; {\rm cm}^{-1}$ and $\sigma \simeq 0.14 \; \mu{\rm m}^{-1} $ \cite{r32}. Assuming a spacing $d= 12\; \mu$m between adjacent  waveguides in the lattice (corresponding to $J \simeq 2.59 \; {\rm cm}^{-1}$) and in the geometrical setting of Fig.4(a) with $h= 25\; \mu$m and $\alpha=\pi/6$, one has $\rho_0 / J \simeq 0.162$, $\rho_1 / J \simeq 0.055$, $\rho_2 / J  \simeq 0.006$, while the higher-order couplings can be neglected. Figures 4(b) and (c) show typical behaviors of the decay dynamics of light intensity $|a(z)|^2$ trapped in waveguide W, excited at the initial plane $z=0$, for two different values of $n_0$ [$n_0$=24 in (b) and $n_0=22$ in (c), corresponding to $\phi=0$  and $\phi= \pi$, respectively], up to a propagation distance $z=10$ cm. Note that, according to the theoretical analysis, in the former case [Fig.4(b)] the decay is accelerated after the propagation distance $\tau = n_0 /(2 J) \simeq 4.26 \; {\rm cm}$, while in the former case the decay is basically suppressed for distances larger than $\tau$.\\
\begin{figure}[htb]
\centerline{\includegraphics[width=8.7cm]{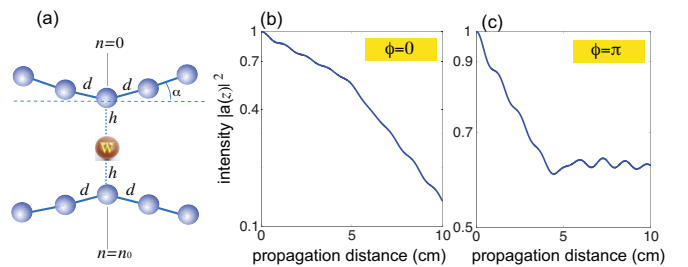}} \caption{ \small
(Color online) Numerical simulations of light intensity decay in a waveguide lattice in physical units. (a) Geometrical setting of the lattice near the coupling region with waveguide W. (b,c) Decay behavior of light intensity $|a(z)|^2$ (on a log scale) versus propagation distance $z$ for (b) $n_0=24$, and (c) $n_0=22$. Other parameter values are given in the text. }
\end{figure} 
{\it Conclusions.} We suggested an integrated optics setup where photon escape from a waveguide non-locally coupled to a waveguide lattice effectively emulates the decay of a giant atom in a 
featureless continuum of bosonic modes. Besides of providing an experimentally accessible 
tool for the simulation of non-Markovian decay of giant atoms, photon escape dynamics in integrated optic settings  could offer a platform for the observation of other exotic effects of light-matter interaction,  such as non-Markovian collective emission from macroscopically-spaced quantum  emitters \cite{r41}.\\
\par

The author declares no conflicts of interest.

\newpage


\clearpage
 {\bf References with full titles}\\
 \\
 \noindent
1.  S. Haroche and J.-M. Raimond, 
{\em Exploring the Quantum: Atoms, Cavities, and Photons} (Oxford University Press, New York, 2006).\\
2. V. Weisskopf and E. Wigner,
Berechnung der naturlichen Linienbreite auf Grund der Diracschen Lichttheorie,
Z. Phy. {\bf 63} 54 (1930).\\
3. A. Frisk Kockum, P. Delsing, and G. Johansson, Designing
frequency-dependent relaxation rates and Lamb
shifts for a giant artificial atom, Phys. Rev. A {\bf 90}, 013837
(2014).\\
4. M. V. Gustafsson, T. Aref, A. F. Kockum, M. K.
Ekstrom, G. Johansson, and P. Delsing, Propagating
phonons coupled to an artificial atom, Science {\bf 346}, 207
(2014).\\
5. L. Guo, A. Grimsmo, A. F. Kockum, M. Pletyukhov, and G. Johansson,
Giant acoustic atom: A single quantum system with
a deterministic time delay, Phys. Rev. A {\bf 95}, 053821 (2017).\\
6. R. Manenti, A. F. Kockum, A. Patterson, T. Behrle,
J. Rahamim, G. Tancredi, F. Nori, and P. J. Leek, Circuit
quantum acousto-dynamics with surface acoustic waves,
Nature Commu. {\bf 8}, 975 (2017).\\
7. A. F. Kockum, G. Johansson, and F. Nori, Decoherence-free
interaction between giant atoms in waveguide quantum
electrodynamics, Phys. Rev. Lett. {\bf 120}, 140404
(2018).\\
8. G. Andersson, B. Suri, L. Guo, T. Aref, and P. Delsing,
Non-exponential decay of a giant artificial atom, Nature Phys. {\bf15}, 1123 (2019).\\
9. A. Gonzalez-Tudela, C. Sanchez Munoz, and J. I. Cirac, Engineering
and Harnessing Giant Atoms in High-Dimensional
Baths: A Proposal for Implementation with Cold Atoms, Phys.
Rev. Lett. {\bf 122}, 203603 (2019).\\
10. B. Kannan, M. Ruckriegel, D. Campbell, A.F. Kockum, J. Braumuller, D. Kim,
M. Kjaergaard, P. Krantz, A. Melville, B.M. Niedzielski, A. Vespalainen, R.
Winik, J. Yoder, F. Nori, T.P. Orlando, S. Gustavsson, and W.D. Oliver, 
Waveguide Quantum Electrodynamics with Giant Superconducting Artificial Atoms, arXiv: 1912.12233v1 (2019).\\
11. L. Guo, A. F. Kockum, F. Marquardt, and G. Johansson, Oscillating
bound states for a giant atom, arXiv:1911.13028 (2019).\\
12. S. Guo, Y. Wang, T. Purdy, and J. Taylor, Beyond Spontaneous Emission: Giant Atom Bounded in Continuum, arXiv:1912.09980v1 (2019).\\
13. A. F. Kockum, Quantum optics with giant atoms-- the first five years, arXiv:1912.13012v1 (2019).\\
14. D. Christodoulides, F. Lederer, and Y. Silberberg, Discretizing light behaviour in linear and nonlinear waveguide lattices, Nature {\bf 424}, 817 (2003).\\
15. S. Longhi, Quantum-optical analogies using photonic structures, Laser Photon. Rev. {\bf 3}, 243 (2009).\\
16. A. Szameit and S. Nolte, Discrete optics in femtosecond-laser-written photonic structures, J. Phys. B {\bf 43}, 163001 (2010).\\
17. I. L. Garanovich, S. Longhi, A. A. Sukhorukov, and Y. S. Kivshar, Light propagation and localization in modulated photonic lattices and waveguides, Phys. Rep. {\bf 518}, 1 (2012).\\
18. A. Aspuru-Guzik and P. Walther, Photonic quantum simulators, Nature Phys. {\bf 8}, 285 (2012).\\
19. S. Longhi, Nonexponential Decay Via Tunneling in Tight-Binding Lattices and the Optical Zeno Effect,
Phys. Rev. Lett. {\bf 97}, 110402 (2006).\\
20. P. Biagioni, G. Della Valle, M. Ornigotti, M. Finazzi, L. Duo, P. Laporta, and S. Longhi,
Experimental demonstration of the optical Zeno effect by scanning tunneling optical microscopy, Opt. Express {\bf 16}, 3762 (2008).\\
21. F. Dreisow, A. Szameit, M. Heinrich, T. Pertsch, S. Nolte, A. Tunnermann, and S. Longhi, Decay Control via Discrete-to-Continuum Coupling Modulation in an Optical Waveguide System,
Phys. Rev. Lett. {\bf 101}, 143602  (2008).\\
22. A. Crespi, F.V. Pepe, P. Facchi, F. Sciarrino, P. Mataloni, H. Nakazato, S. Pascazio, and R. Osellame, Experimental Investigation of Quantum Decay at Short, Intermediate, and Long Times via Integrated Photonics, Phys. Rev. Lett. {\bf 122}, 130401 (2019).\\
23. S. Longhi, Jaynes-Cummings photonic superlattices, Opt. Lett. {\bf 36}, 3407 (2011).\\
24. A. Crespi, S. Longhi, and R. Osellame, Photonic Realization of the Quantum Rabi Model,
Phys. Rev. Lett. {\bf 108}, 163601 (2012).\\
25. B. M. Rodriguez-Lara, F. Soto-Eguibar, A.Z. Cardenas, and H. M. Moya-Cessa, A classical simulation of nonlinear Jaynes-Cummings and Rabi models in photonic lattices, Opt. Express {\bf 21}, 12888 (2013).\\
26. B. M. Rodriguez-Lara, Intensity-dependent quantum Rabi model: spectrum, supersymmetric partner, and optical simulation,
J. Opt. Soc. Am. B {\bf 31}, 1719 (2014).\\
27. S. Longhi, Optical analogue of coherent population trapping via a continuum in optical waveguide arrays, J. Mod. Opt. {\bf 56}, 729 (2009).\\
28. Dreisow, A. Szameit, M. Heinrich, R. Keil, S. Nolte, A. Tunnermann, and S. Longhi, Adiabatic transfer of light via a continuum in optical waveguides, Opt. Lett. {\bf 34}, 2405 (2009).\\
29. A. Crespi, L. Sansoni, G. Della Valle, A. Ciamei, R. Ramponi, F. Sciarrino, P. Mataloni, S. Longhi, and R. Osellame, Particle Statistics Affects Quantum Decay and Fano Interference,
Phys. Rev. Lett. {\bf 114}, 090201 (2015).\\
30. S. Longhi, Quantum simulation of decoherence in optical waveguide lattices, Opt. Lett. {\bf 38}, 4884 (2013).\\
31. R. Keil, C. Noh, A. Rai, S- Stutzer, S. Nolte, D.G. Angelakis, and A. Szameit, Optical simulation of charge conservation violation and Majorana dynamics, Optica {\bf 2}, 454(2015).\\
32. L.J. Maczewsky, K. Wang, A.A. Dovgiy, A.E. Miroshnichenko, A. Moroz, M. Ehrhardt, M. Heinrich, D.N. Christodoulides, A. Szameit, and A.A. Sukhorukov,
Synthesizing multi-dimensional excitation dynamics and localization transition in one-dimensional lattices, Nature Photon. {\bf 14}, 76 (2020).\\
33. A. Peruzzo, M. Lobino, J. C. F. Matthews, N. Matsuda, A. Politi, K. Poulios, X.-Q. Zhou, Y. Lahini, N. Ismail, K.
Worhoff, Y. Bromberg, Y. Silberberg, M. G. Thompson,
and J. L. O$^{\prime}$Brien, Quantum walks of correlated particles, Science {\bf 329}, 1500 (2010).\\
34. G. Corrielli, A.Crespi, G. Della Valle, S. Longhi, and O. Osellame, Fractional Bloch oscillations in photonic lattices,
Nature Commun. {\bf 4}, 1555 (2013).\\
35. S. Longhi, Bound states in the continuum in a single-level Fano-Anderson model, Eur. Phys. J. B {\bf 57}, 45 (2007).\\
36. P Facchi, H Nakazato, and S Pascazio, From the quantum Zeno to the inverse quantum Zeno effect,
Phys. Rev. Lett. {\bf 86}, 2699 (2001).\\
37. R.J. Cook and P.W. Milonni,
Quantum theory of an atom near partially reflecting walls,
Phys. Rev. A {\bf 35}, 5081 (1987).\\
38. T. Tufarelli, F. Ciccarello, and M. S. Kim, Dynamics of spontaneous emission in a single-end photonic waveguide, Phys. Rev. A {\bf 87}, 013820 (2013).\\
39. T. Tufarelli, M.S. Kim, and F. Ciccarello,
Non-Markovianity of a quantum emitter in front of a mirror,
Phys. Rev. A {\bf 90}, 012113 (2014).\\
40. N. Vats and S. John, Non-Markovian quantum fluctuations and superradiance near a photonic band edge, Phys. Rev. A {\bf 58}, 4168 (1998).\\
41. C.W. Hsu, B. Zhen, A.D. Stone, J.D. Joannopoulos, and M.Soljacic, Bound states in the continuum,
Nature Rev. Mat. {\bf 1}, 16048 (2016).\\
42. S. Weimann, Y. Xu, R. Keil, A.E. Miroshnichenko, A. Tunnermann, S. Nolte, A.A. Sukhorukov, A. Szameit, and Y.S. Kivshar, Compact Surface Fano States Embedded in the Continuum of Waveguide Arrays,
Phys. Rev. Lett. {\bf 111}, 240403 (2013).\\
43. K. Sinha, P. Meystre, E.A. Goldschmidt, F.K. Fatemi, S.L. Rolston, and P. Solano, Non-Markovian Collective Emission from Macroscopically Separated Emitters,
Phys. Rev. Lett. {\bf 124}, 043603 (2020).\\

\end{document}